\documentclass[aps, prd, twocolumn, showpacs, superscriptaddress, groupedaddress]{revtex4} 
\usepackage{graphicx}    
\usepackage{amssymb}
\usepackage{dcolumn}
\usepackage{color}
\usepackage{subfigure, rotating, bm, array}
\usepackage[pagebackref=false, colorlinks=true]{hyperref}
\hypersetup{
linkcolor=blue,     
citecolor=blue,     
urlcolor=blue}   
\usepackage{soul}
\usepackage{amsmath}
\newcommand*{\field}[1]{\mathbb{#1}}%
\begin{document}
\title{Strong curvature naked singularities in spherically symmetric perfect fluid collapse          }
\author{Karim Mosani}
\email{kmosani2014@gmail.com}
\affiliation{BITS Pilani K.K. Birla Goa Campus, Sancoale, Goa-403726, India}
\author{Dipanjan Dey}
\email{dipanjandey.adm@charusat.edu.in}
\affiliation{International Center for Cosmology, Charusat University, Anand 388421, Gujarat, India}
\author{Pankaj S. Joshi}
\email{psjprovost@charusat.ac.in}
\affiliation{International Center for Cosmology, Charusat University, Anand 388421, Gujarat, India}

\date{\today}

\begin{abstract}
We investigate here the locally naked singularity formed due to a spherically symmetric inhomogeneous collapsing cloud having non-zero isotropic pressure, in terms of its strength. Sufficient condition provided by Clarke and Krolak for it to be Tipler strong has been used to restrict the parameters 
that represent the non-linear relation between the physical radius and the radial coordinate of the outgoing radial null geodesic at the singular center. Studying end state of a collapsing cloud requires information about the dynamics of collapse, which is unknown in a  general scenario. Hence we study  small perturbations to the mass profile for inhomogeneous dust, which is possible using the formalism developed here. This perturbed mass profile, in turn, gives rise to non-zero pressure. We show the existence of a non-zero measure set of initial data giving rise to such strong curvature naked singularity. 
\\
\\
$\textbf{key words}$: Black Hole, Cosmic Censorship Hypothesis, Gravitational Collapse, Naked Singularity, Strong Singularity.

\end{abstract}
\maketitle

\section{Introduction}
When a sufficiently massive astrophysical body undergoes unhindered gravitational collapse, the end state is an infinitely dense spacetime singularity according to the general theory of relativity.  Such a singularity is, however, always hidden from an outside observer, according to the cosmic censorship hypothesis (CCH) \cite{penrose}. The strong form of CCH, suggesting that no non-spacelike geodesics can escape the singularity, has lately been under scrutiny because it can now be shown that the weak energy condition or the positivity of energy density guarantees a non-zero measured set of such geodesics. Formation of such locally naked singularity due to the gravitational collapse of a Lemaitre-Tolman-Bondi (LTB) dust cloud, which has zero pressure, was shown to be possible by Joshi and Dwivedi \cite{joshi}, under generic initial conditions. The significant role played by the inhomogeneity of the collapsing cloud was highlighted in such a phenomena. However, unlike the LTB dust cloud, a more realistic star does have non-zero pressure. Additionally, the matter is expected to behave like a perfect fluid at the center of the cloud as discussed in \cite{goswami}. The formalism developed to investigate the end state of a collapsing cloud having non-zero pressure suggests that the local nakedness or otherwise of the singularity thus formed, depends upon the polarity of the smallest non-zero component of the Taylor expansion of the singularity curve \cite{joshi2,joshi3}. The positivity of such component implies that the tangent of the outgoing radial null geodesic (ORNG) is positive at the singularity, implying that the singularity is at least locally naked. 

Such locally naked singularities, however, may not be considered as evidence for the defiance of cosmic censorship if they are gravitationally weak. Any object hitting the singularity, if crushed to zero volume, is called ``strong" curvature singularity according to Tipler \cite{tipler}.  It was shown by Newman \cite{newman} that naked singularities investigated by Eardley and Smarr \cite{eardley} and Christodoulou \cite{christodoulou}, formed in the classes of LTB cloud collapse are weak. Using the sufficient condition given by Clarke and Krolak \cite{clarke} for a singularity to be strong in the sense of Tipler, naked singularity formation due to collapsing self-similar marginally bound singularity was studied by Waugh and Lake \cite{waugh} and independently by Ori and Piran \cite{ori}. Later, Joshi and Dwivedi investigated the naked singularity formed due to the collapse of LTB dust cloud under generic initial conditions for collapse, and derived the value of a certain parameter $\alpha$ for which the singularity is gravitationally strong \cite{joshi}. The parameter $\alpha$ physically signifies the non-linear relation between the physical radius of the cloud and the radial coordinate of the ORNG at the singular center ($\alpha=1$ corresponds to a linear relation). The stability of such singularities against some perturbations in the initial data was later shown by Deshingkar, Joshi and Dwivedi \cite{deshingkar}. Coming to the collapsing cloud having non-zero pressure, many models have been studied in which naked singularities are shown to arise (see e.g. \cite{ori1}-\cite{goncalves} and \cite{joshi4} for a review). Here, we study the singularities formed due to a collapsing spherical cloud made up of a perfect fluid with non-zero pressure and derive an analogous criterion needed to be imposed on $\alpha$ for the singularity to be strong. Our basic purpose here is thus to examine and characterize the conditions that ensure that the naked singularities forming in collapse with non-zero pressure are strong curvature in nature.  

The paper is arranged as follows: Einstein's field equations, regularity conditions, and the mathematical formalism to understand the final state of a collapsing spherically symmetric perfect fluid with arbitrary pressure are discussed in Section II. The strength of singularity and related results are discussed in Section III. An example showing the existence of a non-zero measured set of initial data giving rise to locally naked Tipler strong singularity is then illustrated and worked out in Section IV. Concluding remarks are given in Section V. 

\section{Collapse formalism}
The collapse of a spherically symmetric cloud made up of perfect fluid is governed by three functions $\nu(t,r)$, $\psi (t,r)$ and $R(t,r)$, and the metric is expressed as:
\begin{equation} \label{metric}
    ds^2=-e^{2\nu (t,r)}dt^2+e^{2\psi(t,r)}dr^2+R^2(t,r)d\Omega ^2
\end{equation}
in the comoving coordinates $t$ and $r$. The stress-energy tensor for a general type I matter field, more specifically a perfect fluid, has non-diagonal terms as zero, and diagonal terms as
\begin{equation}
    T^t_t=-\rho, T^r_r=T^{\theta}_{\theta}=T^{\phi}_{\phi}=p.
\end{equation} 
Here the $\rho$ is the energy density and $p$ is the isotropic pressure of the collapsing cloud. The matter field under consideration is assumed to be satisfying the weak energy condition thereby restricting the components of stress-energy tensor in the following way:
\begin{equation}
 \rho\geq 0, \rho+p\geq 0.   
\end{equation}
In the units of $8\pi G=c=1$, the Einstien's field equations relates the metric functions $\nu(t,r)$, $\psi (t,r)$ and $R(t,r)$ with the components of stress-energy tensor in the following way:
\begin{eqnarray}
      \rho &=\frac{F'}{R^2R'},\label{efe1} \\
     p &=-\frac{\dot F}{R^2\dot R}, \label{efe2}\\
     \nu ' &=-\frac{p'}{\rho +p}, \label{efe3}\\
     2 \dot R' &= R'\frac{\dot G}{G}+ \dot R\frac{H'}{H},\label{efe4}
\end{eqnarray}
where,
\begin{equation} \label{gh}
    G(t,r)=e^{-2\psi}R'^2; \hspace{0.5cm} H(t,r)=e^{-2 \nu}\dot R^2.
\end{equation}
The superscript dot and prime are the notations used for partial derivative with respect to time and radial coordinates respectively. Here $F$ is the Misner-Sharp mass function given by
\begin{equation}\label{msmf}
     F=R(1-G+H).
\end{equation}
It physically signifies the mass of the cloud inside a shell of radius $r$ at time $t$. It can be expressed as $F=r^3\mathcal{M}$, where $\mathcal{M}$ is such that it maintains regularity. By regularity, we mean that $\mathcal{M}$ is a suitably differentiable function which does not blow up or vanish as $r \to 0$. Doing so ensures that the energy density at the regular center does not blow up before the formation of central shell focusing singularity. Another regularity condition needed to be fulfilled by the collapsing matter field to be well behaved is the absence of cusp in the energy density at the center which is taken care by the equation
\begin{equation}
    \mathcal{M}'(t,0)=0.
\end{equation}
The physical radius of the cloud is represented by the component of metric, $R(t,r)$. For different shells to avoid crossing each other, $R$ has to follow the inequality $R'>0$. To get a collapsing solution of Einstein's field equations, we have to restrict $\dot R (t,r)$ to be less than zero. This indicates that given a shell of radial coordinate $r$, the corresponding physical radius $R$ decreases as time passes until it becomes a singularity, i.e. $R(t,r)=0$. It is to be noted that $R(t,r)$ vanishes also at the regular center, i.e. at $r=0$. This means that vanishing $R(t,r)$ does not necessarily imply the formation of a singularity. The representation of the distinction in both the cases can be achieved by expressing $R$ as
\begin{equation}\label{R}
    R(t,r)=r v(t,r),
\end{equation}
where $v(t,r)$ can be viewed as a scale factor. The scaling freedom accessible for $r$ can be used to define $R(t_i,r)=r$, where $t_i$ is the initial epoch. This allows us to write the following:
\begin{equation}
   v(t_i,r)=1; \hspace{0.5cm} v(t_s(r),r)=0; \hspace{0.5cm} \dot v<0,
\end{equation}
where $t_s(r)$ is called the singularity curve which gives the time of formation of singularity due to collapsing shell having radial coordinate $r$. Now it can be said that this shell collapses to form a singularity if $v(t_s,r)=0$, thereby distinguishing the case from a regular center. An additional benefit of introducing the scale factor $v$, also known alternatively as the scaling function, is the freedom to study the collapse formalism in the transformed $(r,v)$ coordinates, instead of $(t,r)$ coordinates, which will be apparent in the forthcoming approach.

Let us now recall briefly the formalism developed earlier \cite{joshi2, joshi3} to study the end state of the collapse. We start with defining an appropriately differentiable function $A(r,v)$ as follows:
\begin{equation}\label{av}
    A_{,v}=\nu ' \frac{r}{R'}.
\end{equation}
Eq.(\ref{efe3}), after integrating, can be used to express $G$ in terms of $A(r,v)$ as
\begin{equation} \label{G2}
    G(r,v)=b(r)e^{2A(r,v)},
\end{equation}
where the integration constant $b(r)$ is related to the velocity with which the matter shell falls in. It can be expressed near the regular center as
\begin{equation}
    b(r)=1+r^2b_0(r).
\end{equation}
$b_0(r)$ is interpreted in analogy with the Lemaitre Tolman Bondi (LTB) dust model in which $b_0<0$ means bounded, $b_0>0$ means unbounded and $b_0=0$ means marginally bound dust collapse. The equation of motion can be found using Eq.(\ref{msmf}) as 
\begin{equation} \label{eom}
    \sqrt{v}\dot v=-e^{\nu} \sqrt{\mathcal{M}+\frac{v(be^{2A}-1)}{r^2}}.
\end{equation}
This can be integrated to achieve the time curve $t(r,v)$ as follows:
\begin{equation}\label{timecurve}
    t(r,v)=t_i+\int_v^1 \frac{e^{-\nu}}{\sqrt{\frac{\mathcal{M}}{v}+\frac{be^{2A}-1}{r^2}}}dv.
\end{equation}
The time curve dictates the time required for a collapsing shell of radial coordinate $r$ to arrive at an event $v$. This could now be used to get the singularity curve,
\begin{equation}\label{singularitycurve}
    t_s(r)=t(r,0)=t_i+\int_0^1 \frac{e^{-\nu}}{\sqrt{\frac{\mathcal{M}}{v}+\frac{be^{2A}-1}{r^2}}}dv
\end{equation}
which tells us the time required for a shell of radial coordinate $r$ to collapse to a singularity. Near the center, the time curve can be Taylor expanded  around $r=0$ as 
\begin{equation}\label{taylorsinglaritycurve}
    t(r,v)=t(0,v)+r\chi_1(v)+r^2\chi_2(v)+r^3\chi_3(v)+O(r^4),
\end{equation}
where
\begin{equation}\label{chii}
    \chi_i(v)=\frac{1}{i!}\frac{d^i t}{dr^i}\bigg |_{r=0}.
\end{equation}
For a singularity to be at least locally naked, there have to be families of timelike or null geodesics leaving the singularity. If the trapped surfaces in the neighborhood around the center are formed before the formation of the singularity, the geodesics will not be able to escape, thereby giving a black hole as the end product. The existence or otherwise of such escaping  geodesics can be investigated by considering the equation for outgoing radial null geodesics (ORNG) as follows: 
\begin{equation}\label{efong}
    \frac{dt}{dr}=e^{\psi-\nu}.
\end{equation}
If these geodesics were to be incomplete in the past at the singularity, $R \to 0$ as $t \to t_s$ (or $v\to 0$) along these curves, that ensures a visible singularity. The above equation can be expressed using chain rule in terms of $R$ and $u=r^{\alpha}$, where $\alpha>1$,  as
\begin{equation}
    \frac{dR}{du}=\frac{1}{\alpha}\frac{R'}{r^{\alpha-1}}\left (1+ \frac{\dot R}{R'}e^{\psi-\nu} \right ). 
\end{equation}
which can be rewritten as
\begin{equation} \label{dRbydu}
    \frac{dR}{du}=\frac{1}{\alpha} \left ( \frac{R}{u}+ \frac{\sqrt{v}v'r^{\frac{5-3\alpha}{2}}}{\sqrt{\frac{R}{u}}} \right ) \left (\frac{1-\frac{F}{R}}{\sqrt{G}(\sqrt{G}+\sqrt{H})} \right ).
\end{equation}
Here we have used the relation obtained from Eq.(\ref{msmf}). Along constant $v$ surface, $dv=v'dr+\dot v dt=0$, and hence, $\sqrt{v}v'$, appearing in the above equation, could be obtained from Eq.(\ref{eom})as
\begin{equation}\label{sqrtv'}
    \sqrt{v}v'=e^{\psi}\sqrt{e^{2A}vb_0+vh+\mathcal{M}},
\end{equation}
where
\begin{equation}
    h(r,v)=\frac{e^{2A}-1}{r^2}.
\end{equation}
For a singularity to be naked (at least locally), the tangent to the future directed radially null geodesic, which ceases at the singularity in the past, should have $\frac{dR}{du}>0$ at the singularity in the $(R,u)$ plane \cite{joshi2}. Also, it should be finite. L'Hospital's rule then gives us
\begin{equation} \label{x0}
    X_0= \lim_{(R,u)\to (0,0)} \frac{R}{u}=\frac{dR}{du}.
\end{equation}
The mass profile $\mathcal{M}$ near the center can be Taylor expanded around $r=0$ as
\begin{equation}\label{temp}
    \mathcal{M}(r,v)=M_0(v)+M_2(v)r^2+M_3(v)r^3+M_4(v)r^4+o(r^5).
\end{equation}
At the limit $(r,v)\to (0,0)$ we obtain
\begin{equation}\label{limsqrtvv'}
\begin{split}
\lim_{(r,v)\to 0}\sqrt{v}v'= & \big (\chi_1(0) +2r\chi_2(0)+3r^2\chi_3(0) \\
& +4r^3\chi_4(0)+ o(r^4)\big )  \sqrt{M_0(0)}.
\end{split}
\end{equation}
Substituting for $\sqrt{v}v'$ from Eq.(\ref{limsqrtvv'}) in the limiting case of Eq.(\ref{dRbydu}) along with using Eq.(\ref{taylorsinglaritycurve}-\ref{chii})  and Eq.(\ref{x0}) gives
\begin{equation}\label{X02}
\begin{split}
X_{0}^{\frac{3}{2}}= & \lim_{r\to 0}\frac{1}{\alpha-1}\big ( \chi_1(0) +2r\chi_2(0)+3r^2\chi_3(0) \\
& +4r^3\chi_4(0)+o(r^4)\big ) \sqrt{M_0(0)}r^{\frac{5-3\alpha}{2}}.
\end{split}
\end{equation}
It can be seen from the above equation that the problem of determining the local nakedness of the singularity is reduced to determining the polarity of $X_0$. Eq.(\ref{X02}) depicts the relation between the tangent of ORNG at singularity $X_0$ and the components $\chi_i$ of the Taylor expansion of the singularity curve. Here, a specific value of $\alpha$ is chosen so that $X_0 \neq 0$. For instance, if $\chi_1 \neq 0$ then $\alpha=5/3$ has to be chosen, and Eq.(\ref{X02}) is reduced to
\begin{equation}
    X_0^{\frac{3}{2}}=\frac{3}{2}\chi_1(0)\sqrt{M_0(0)}
\end{equation}
at the limit $r\to 0$. This implies that polarity of $\chi_1(0)$ is  the deciding factor for the local visibility or otherwise of the singularity. 

Another possible value which $\alpha$ can take is $\alpha=7/3$, for which the deciding factor is $\chi_2$ as seen in the following specific form of Eq.(\ref{X02}) as follows:
\begin{equation}
    X_0^{\frac{3}{2}}=\frac{3}{2}\chi_2(0)\sqrt{M_0(0)}.
\end{equation}
Here, $\chi_1(0)$ should be of the order of $r$ in order to avoid the blowing up of $X_0$, hence $\chi_1(0)$ has to be zero in the limit $r\to 0$. Generally, $\alpha$ is restricted to the following values so that $X_0 \neq 0$:
\begin{equation}\label{alpha}
  \alpha \in \left\{\frac{2n+3}{3};\hspace{0.2cm} n\in \field{N} \right\}.
\end{equation}
Additionally, near $(r,v)\to (0,0)$,  we should have 
\begin{equation}\label{orderofchi}
  \chi_i(v) \sim O\left(r^{\frac{3\alpha-1}{2}-i}\right), \hspace{0.5cm} \forall \ i<\frac{3}{2}(\alpha-1).
\end{equation}
This ensures that $\chi_i(0)=0$ and thereby preventing $X_0$ from blowing up. Whether or not these values of $\alpha$ in (\ref{alpha}) corresponds to a singularity which is strong, in the sense of Tipler, is investigated in the next section.

\section{Strength of Singularities}
The tangents of the outgoing timelike or null geodesic from a singularity formed due to gravitational collapse of an inhomogeneous spherically symmetric perfect fluid with non-zero pressure are as follows:

\begin{equation} \label{tangent equation}
    \begin{split}
& K^t =\frac{dt}{d\lambda}=\frac{\mathcal{P}}{R}, \\ 
& K^r =\frac{dr}{d\lambda}=\frac{\sqrt{G}}{RR'}\sqrt{\mathcal{P}^2\frac{\dot R^2}{H}-l^2+BR^2},\\ 
& K^{\theta^2}+\sin ^2\theta K^{\phi^2} =\frac{l^2}{R^4}.\\
\end{split}
\end{equation}

Here the value of $B$ denotes the type of geodesics such that for null geodesic $B=0$ and for timelike geodesic $B=-1$. Also, $l$ is called the impact parameter which vanishes for radial geodesics. The function $\mathcal{P}(t,r)$ satisfies the following geodesic equation:
\begin{widetext}
\begin{equation}\label{geodesic equation}
\begin{split}
& \frac{d\mathcal{P}}{d\lambda}-\frac{\mathcal{P}^2}{R}\left(\frac{\dot R}{R} -\frac{ \dot H}{2H}+\frac{\ddot R}{\dot R} -\frac{\dot R'}{R'}+\frac{\dot G}{2G} \right)-\frac{\mathcal{P}\sqrt{G}}{R}\sqrt{\frac{\mathcal{P}^2\dot R^2}{H}-l^2+BR^2} \left (\frac{1}{R}+\frac{1}{R'}\left( \frac{2\dot R'}{\dot R}-\frac{H'}{H} \right)\right) \\
& +\frac{H}{\dot R}\left(\frac{l^2}{R}\left(\frac{1}{R}+\frac{\dot G}{2G \dot R}-\frac{\dot R'}{\dot R R'}\right)+\frac{BR}{\dot R}\left(\frac{ \dot R'}{R'}-\frac{\dot G }{2G}\right)\right)=0.
\end{split}
\end{equation}
\end{widetext}
For radial null geodesic, close to $\lambda=0$, i.e. near the singularity, using L'Hospital's rule in the above equation gives us the expression of $\mathcal{P}$ as follows:
\begin{equation} \label{P expression}
\begin{split}
    \mathcal{P}=& \lim_{r\to 0}\frac{R}{\lambda}\Bigg (\frac{\dot R}{R}-\frac{\dot H}{2H}+\frac{\ddot R}{\dot R}+\frac{\dot G}{2G}-\frac{\dot R'}{R'}+\sqrt{\frac{G}{H}}\Bigg (\frac{\dot R}{R} \\
    & -\frac{H'\dot R}{HR'}+\frac{2\dot R'}{R'}\Bigg )\Bigg )^{-1}.
    \end{split}
\end{equation}
The sufficient condition for a singularity to be strong in the sense of Tipler \cite{tipler}, provided by Clarke and Krolak \cite{clarke}, is that at least along one null geodesic with the affine parameter $\lambda$, with $\lambda=0$ at the singularity, the following inequality should be satisfied:
\begin{equation} \label{Krolak and Clarke criteria}
    \lim_{ \lambda \to 0} \lambda ^2R_{ij}K^iK^j>0.
\end{equation}
Eq.(\ref{efe1}) and Eq.(\ref{efe2}), gives
\begin{equation}\label{psi}
\begin{split}
R_{ij}K^iK^j = & \frac{1}{2R^2}\Big ( \frac{\dot R}{H R'}\left(F'\dot R-3\dot FR'\right)(K^t)^2 \\
    & +\frac{R'}{G\dot R}\left(F'\dot R+\dot FR'\right)(K^r)^2 \Big ).
\end{split}    
\end{equation}
Substituting for the tangents to the radial null geodesic from Eq.(\ref{tangent equation}), we get
\begin{equation}\label{clarkkrolak2}
     \lim_{ \lambda \to 0}\lambda^2R_{ij}K^iK^j = 3 \lim_{ \lambda \to 0}\left(\frac{\lambda \sqrt{F}\mathcal{P}\dot R}{R^2\sqrt{rR'}\sqrt{H}}\right)^2.
\end{equation}
Here, we have used the following limiting values arising from the regularity conditions:
\begin{equation}\label{etazeta}
 \lim_{r\to 0}\frac{rF_{,r}}{F}=3,\hspace{0.5cm} \lim_{v\to 0}\frac{vF_{,v}}{F}=0.
\end{equation}
The particular case of LTB collapse reduces the expression Eq.(\ref{clarkkrolak2}) to 
\begin{equation}
      \lim_{ \lambda \to 0}\lambda^2R_{ij}K^iK^j = 3 \lim_{ \lambda \to 0}\left(\frac{\lambda \sqrt{F}\mathcal{P}}{R^2\sqrt{rR'}}\right)^2,
\end{equation}
which agrees with the result obtained in \cite{joshi}. The above equation is obtained by substituting $H=\dot R^2$ in Eq.(\ref{clarkkrolak2}).
Using Eq.(\ref{P expression}) and Eq.(\ref{clarkkrolak2}), the Clarke and Krolak's criteria is restated as
\begin{equation} \label{clarkkrolak}
\begin{split}
    & \lim_{(r,v)\to (0,0)} \left(\frac{F'}{R'}-\frac{\dot F}{\dot R}\right)\Bigg ( \sqrt{G}\left(1-\frac{H'R}{HR'}+\frac{2R\dot R'}{\dot R R'} \right) \\
    & +\sqrt{H}\left(1-\frac{\dot H R}{2H \dot R}+\frac{\ddot R R }{\dot R^2}+\frac{R\dot G}{2\dot R G}- \frac{R\dot R'}{\dot R R'}\right)\Bigg )^{-2}>0,
\end{split}    
\end{equation}
which should hold at least along one null geodesic which is past incomplete at the singularity, for the singularity to be Tipler strong. $H$ can be expressed using Eq.(\ref{gh}) and Eq.(\ref{eom}) as
\begin{equation}\label{H}
    H(r,v)=\frac{\mathcal{M}r^2}{v}+be^{2A}-1.
\end{equation}
Differentiating Eq.(\ref{H}) with respect to $r$ can lead to 
\begin{equation}\label{H'}
    \lim_{(r,v)\to(0,0)}\frac{H'}{H}=\lim_{r\to 0}\frac{1}{r}+\frac{M_{,r}}{M}.
\end{equation}
Differentiating $H$ in Eq.(\ref{gh}) with respect to $t$ gives the following equation:
\begin{equation}\label{hdot}
    \frac{\ddot R R}{\dot R^2}-\frac{\dot H R}{2H\dot R}=\nu_{,v}v.
\end{equation}
Differentiating $G$ in Eq.(\ref{gh}) with respect to $t$ and using Eq.(\ref{G2}) gives the following equation:
\begin{equation}\label{gdot}
    \frac{r\dot v'}{\dot v}=2vA_{2,v}r^2+2\psi_{,v}v.
\end{equation}
In the $(r,v)$ coordinate, we have
\begin{equation}\label{F'/R'-Fdot/Rdot}
\begin{split}
  \lim_{(t,r)\to(t_s,0)}\frac{F'}{R'}-\frac{\dot F}{\dot R} & =  \lim_{(r,v)\to(0,0)}\frac{1}{2}\left(\frac{F_{,r}}{v}-\frac{F_{,v}}{r} \right) \\
  & =\frac{3}{2}\frac{\mathcal{M}(0,0)}{X_0} \lim_{r\to 0}  r^{3-\alpha}.
 \end{split} 
\end{equation}
Using Eq.(\ref{H'}, \ref{hdot}, \ref{gdot}, \ref{F'/R'-Fdot/Rdot}) in Eq.(\ref{clarkkrolak}) we obtain the condition of Clarke and Krolak as
\begin{equation} \label{final clarke and krolak criteria}
\begin{split}
     \lim_{(r,v)\to (0,0)} & \Bigg (\sqrt{\frac{\lvert X_0 \rvert}{\mathcal{M}(0,0)} }r^{\left(\frac{\alpha-3}{2}\right)}\bigg (\frac{1}{2}-\frac{M_{,r}r}{2M}+2\psi_{,v}v \\
     & + 2vA_{2,v}r^2\bigg ) +1 +\nu_{,v}v-\psi_{,v}v \Bigg )^{-2}>0
\end{split}
\end{equation}
at least along one null geodesic. The above inequality can be satisfied only if 
\begin{equation}\label{alphageq3}
    \alpha\geq 3,
\end{equation}
for if $\alpha<3$, then the denominator on the left hand side of the inequality (\ref{final clarke and krolak criteria}) will blow up in the limit $(r,v)\to (0,0)$, thereby not satisfying the inequality anymore.

From Eq.(\ref{orderofchi}) and Eq.(\ref{alphageq3}) it can be seen that in order to maintain the finiteness of $X_0$, $\chi_1$ and $\chi _2$ should be of the order of at least $r^2$ and $r$ respectively, implying that 
\begin{equation}
\chi_1(0)=\chi_2(0)=0.
\end{equation}

It is to be noted that $\alpha$ can take values as follows:
\begin{equation}
    \alpha \in \left\{\frac{2n+1}{3};\hspace{0.2cm} n\geq 4; \hspace{0.2cm} n\in \field{N} \right\}.
\end{equation}
We now carry out a case study for one such value of $\alpha$  in the next section.

\section{Collapse Endstates}
If $\alpha=3$, the equation for tangent of the null geodesic at the singularity for $r=0$ follows from Eq.(\ref{X02}) as 
\begin{equation}\label{X0chi3}
      X_{0}^{\frac{3}{2}}=\lim_{r\to 0}\frac{3}{2}\sqrt{M_0(0)}\chi_3(0).
\end{equation}
Polarity of $\chi_3$ then determines the polarity of $X_0$ which in turn determines the nakedness or otherwise of the Tipler strong singularity. Substituting for density and pressure of the cloud from Eq.(\ref{efe1}) and Eq.(\ref{efe2}) in Eq.(\ref{efe3}) gives us 
\begin{equation}\label{nudash}
    \nu '=\frac{\mathcal{M},_{vr}v+\left(\mathcal{M},_{vv}v-2\mathcal{M},_{v} \right)w}{\left(3\mathcal{M}+r\mathcal{M},_r-\mathcal{M}_v v\right)v}R'.
\end{equation}
Here, $v'$, which is the partial derivative of $v$ in $(t,r)$ coordinate has been expressed as a function $w(r,v)$ in the $(r,v)$ coordinate.  One could use the above equation in Eq.(\ref{av}) for obtaining the integral expression of $A(r,v)$ as
\begin{equation}\label{iearv}
    A(r,v)=\int_v^1 \frac{\mathcal{M},_{vr}v+\left(\mathcal{M},_{vv}v-2\mathcal{M},_{v} \right)w}{\left(3\mathcal{M}+r\mathcal{M},_r-\mathcal{M}_v v\right)v} rdv.
\end{equation}
Also regularity condition demand that $A \simeq r^2 $. Hence one can Taylor expand it around $r=0$ as 
\begin{equation}\label{tearv}
    A(r,v)=A_2(v)r^2+A_3(v)r^3+...
\end{equation}
where the components $A_i(v)$, $i\geq 2$, can be obtained using Eq.(\ref{iearv}) as follows:
\begin{equation}\label{A2}
    A_2(v)=\int_v^1 \frac{2M_{2,v}+\left(M_{0,vv}-\frac{2M_{0,v}}{v} \right)w,_{r}}{3M_0-M_{0,v}v} dv,
\end{equation}
\begin{equation}\label{A3}
    A_3(v)=\int_v^1 \frac{6M_{3,v}+\left(M_{0,vv}-\frac{2M_{0,v}}{v} \right)w,_{rr}}{3M_0-M_{0,v}v}dv,
\end{equation}
\begin{widetext}
\begin{equation}\label{A4}
    \begin{split}
 A_4(v)= & \int_v^1\frac{1}{\left(3M_0-M_{0,v}v\right)^2}\Bigg (2M_{2,v}\left(vM_{2,v}-5M_2\right)+4M_{4,v}\left(3M_{0}-vM_{0,v}\right) + w,_r \bigg ( M_0\big (3M_{2,vv} \\ &-\frac{6M_{2,v}}{v}\big )+M_{0,v}\left(\frac{10M_2}{v}-vM_{2,vv}\right)+M_{0,vv}\left(v M_{2,v}-5M_2\right)\bigg )\Bigg ) +w,_{rrr}\frac{\left(M_{0,vv}v-2M_{0,v}\right)}{6v\left(3M_0-M_{0,v}v\right)} dv,
    \end{split}
\end{equation}
\end{widetext}
and
\begin{widetext}
\begin{equation}\label{A5}
    \begin{split}
A_5(v)= & \int_v^1\frac{1}{(3M_0-M_{0,v}v)^2}
\Bigg (-12M_3M_{2,v}-15M_2M_{3,v}+5v M_{2,v}M_{3,v} + w,_{r} \Big (-\frac{6M_0M_{3,v}}{v} \\
& +M_3\left(\frac{12M_{0,v}}{v}-6M_{0,vv}\right)+M_{0,vv}M_{3,v}v+3M_0M_{3,vv}-M_{0,v}M_{3,vv}v \Big ) +w,_{rr}\Bigg (5M_2\left(\frac{M_{0,v}}{v}-\frac{M_{0,vv}}{2}\right) \\
& +3M_0\left(-\frac{M_{2,v}}{v}+\frac{M_{2,vv}}{2}\right)+\frac{v}{2}\left(M_{2,v}M_{0,vv}-M_{0,v}M_{2,vv}\right)\Bigg )\Bigg) + w,_{rrrr} \frac{\left(-\frac{M_{0,v}}{12v}+\frac{M_{0,vv}}{24}\right)}{3M_0-M_{0,v}v} dv.    \end{split}
\end{equation}
\end{widetext}
These components of Taylor expansion of $A(r,v)$ around $r=0$ are then used to determine $\chi_3$ by differentiating the singularity curve thrice. We also have,
\begin{equation}\label{timecurvecomponent}
    \frac{be^{2A}-1}{r^2}=\sum_{i=0}^{\infty} ((i+2)A_{i+2}+b_{0i})r^i
\end{equation}
 near the center. Here, $b_{0i}$ are the coefficients of $r^i$ in the Taylor expansion of $b_0(r)$ around $r=0$. Substituting from Eq.(\ref{timecurvecomponent}) in Eq.(\ref{singularitycurve}) along with using Eq.(\ref{taylorsinglaritycurve}) and Eq.(\ref{chii}), we obtain the expression of $\chi_3$ as
\begin{widetext}
\begin{equation}\label{chi3}
\begin{split}
    \chi_3=  \int_v^1 \frac{3A_3+b_{01}}{\left(\frac{M_0}{v}+2A_2+b_{00}\right)^{\frac{3}{2}}}\left(\frac{g_2}{2}-\frac{5}{16}\left(\frac{3A_3+b_{01}}{\frac{M_0}{v}+2A_2+b_{00}}\right)^2+\frac{3}{4}\left(\frac{\frac{M_2}{v}+4A_4+b_{02}}{\frac{M_0}{v}+2A_2+b_{00}}\right)\right)- \frac{1}{2}\frac{\left(\frac{M_3}{v}+5A_5+b_{03}\right)}{\left(\frac{M_0}{v}+2A_2+b_{00}\right)^{\frac{3}{2}}}dv.
\end{split}
\end{equation}
\end{widetext}
Here, $g_2=\frac{1}{2}A_{2,v}v.$
As is apparent from Eq.(\ref{X0chi3}), polarity of $\chi_3$ is the deciding factor for local visibility of Tipler strong singularity. The expressions for $A_2$, $A_3$, $A_4$ and $A_5$ can be obtained from Eq.(\ref{A2}-\ref{A5}) for a given mass profile $\mathcal{M}(r,v)$, which is then substituted in Eq.(\ref{chi3}). However, while calculating the $A_i$s, we also require the derivatives of $w(r,v)$ with respect to $r$, which is not known in general. Nevertheless, for a well-chosen mass profile such that the components non-minimally coupled with the derivatives of $w(r,v)$ in the integral expressions for $A_i$ vanish, we could bypass the requirement of the information of the collapse dynamics. Since, there is no mention of equation of state here, we have total five field equations in six unknown parameters namely $p$, $\rho$, $\nu$, $\psi$, $R$ and $F$, i.e. two matter variables, three metric tensor components and the Misner-Sharp mass function. Therefore, there is one degree of freedom left, thereby allowing us to specify the evolution of mass profile $\mathcal{M}$. The idea is to give a small perturbation to the mass profile corresponding to inhomogeneous dust which upto fourth order is expressed close to the center as 
\begin{equation}\label{massprofile}
    \mathcal{M}(r,v)=m_0+m_2r^2+m_3r^3+m_4r^4,
\end{equation}
where $m_0$, $m_2$, $m_3$ and $m_4$ are constants. The perturbation term $\delta(v)$ is then coupled minimally to the fourth order component of $\mathcal{M}$. The reason for this form of perturbation is to vanish the terms involving the derivative of $w(r,v)$ in the expression of $A_i$.
One such example of a perturbed mass profile is as follows:
\begin{equation}\label{perturbedmassprofile}
   \mathcal{M}(r,v)=m_0+m_2r^2+m_3r^3+m_4r^4+\delta(v) r^4.
\end{equation}
This mass profile can give rise to non-zero pressure near the center. 

Now, let us consider the mass profile Eq.(\ref{massprofile}), with  $m_0=1$, $m_2=-0.1$, $m_3=0$ and $m_4=-0.1$. Let us give a fourth order perturbation, $\delta (v)=-0.1 (1-v^2)$. This perturbed mass profile corresponds to a perfect fluid with non-zero pressure associated with it. Fixing $b_{00}=-0.5$ and $b_{01}=-0.1$, a non-zero measured set of initial data $(b_{02},b_{03})$ satisfying the inequality 
$$
9.46857 b_{02}+48.4614 b_{03}<1
$$
is obtained for which $\chi_3>0$, and hence the end state of the collapse for such initial data is a locally visible Tipler strong singularity.

\section{Concluding Remarks}
Some concluding points and open concerns are discussed below:
\begin{enumerate}

\item The necessary criterion for a central shell-focusing singularity formed due to gravitational collapse of a spherically symmetric inhomogeneous perfect fluid with non-zero pressure to be visible is that the relation between the physical radius and the radial coordinate of ORNG should be of the form
\begin{equation*}
        R=X_0r^{\alpha}, \hspace{0.5cm} X_0>0,
    \end{equation*}
where $\alpha$ is restricted to the values given by the set 
\begin{equation*}
    \alpha \in \left\{\frac{2n+3}{3};\hspace{0.2cm} n\in \field{N} \right\}.
\end{equation*}

\item For this singularity to be strong in the sense of Tipler, set of possible values of $\alpha$ is further refined as follows:
\begin{equation*}
     \alpha \in \left\{\frac{2n+3}{3};\hspace{0.2cm} n\geq 3; \hspace{0.2cm} n\in \field{N} \right\}.
\end{equation*}
 This restriction on $\alpha$ concludes that the locally naked singularities in \cite{joshi5} and \cite{mosani} are not Tipler strong because of the fact that $\alpha$ was chosen to be $\frac{5}{3}$ and $\frac{7}{3}$. 
 
\item While investigating the end state, the requirement of pre-knowledge of the dynamics of the collapse, $v(t,r)$, causes a hindrance to proceed further to determine the visibility of the singularity, as observed in Eq.(\ref{A2}-\ref{chi3}). Nevertheless, due to a degree of freedom available with us, we have freedom of choice of fixing an unknown function. In our case, this unknown function is the mass profile of the fluid. By wisely choosing the mass profile, the requirement of the knowledge of  $v(t,r)$ could be bypassed. To achieve this, we have given a perturbation to the mass profile for dust in such a way that the components non-minimally coupled with the derivative terms of the scaling function in Eq.(\ref{A2}-\ref{A5}) vanish. One way to obtain such a mass profile is to add a perturbed term of order four in $r$. For an example of such a form of a mass profile, there indeed exists a non-zero measured set in the $(b_{02},b_{03})$ plane for which the end state after the collapse is a Tipler strong locally visible singularity. Existence of such a set of initial data guarantees that the naked singularities forming due to perfect fluid collapse are stable against any perturbation in the initial data from which the collapse begins.  

\item This acts as a counter-example to at least the strong cosmic censorship hypothesis which does not allow the existence of such locally visible singularity. It is to be noted that the matter fluid formed due to such perturbed term satisfies the weak energy condition and has non-zero pressure $p=-\frac{\delta,_{v}}{X_0^2}$, not restricted to any equation of state. Non-zero pressure in the collapsing cloud arises because of the time dependence property of the perturbed mass profile. Hence, we have shown that there exists collapsing cloud having certain mass profile with non-zero pressure which collapses to form a Tipler strong singularity which is locally visible. Also, since the collapsing cloud is scale independent, if its size is very large, an observer sufficiently close to the singularity will be able to detect the singularity even if it is only locally naked. Hence, even a locally naked singularity is a serious defiance of the cosmic censorship.

\item In \cite{joshi}, in the case of inhomogeneous collapsing dust, it has been shown that $\alpha\leq 3 $ for a singularity to be naked ($X_0>0$). This puts a further restriction on $\alpha$, fixing it to $\alpha=3$ for a singularity to be Tipler strong and locally visible. In our case study, we have shown that $\alpha=3$ indeed gives Tipler strong locally visible singularity formed due to collapsing perfect fluid cloud with non-zero pressure. Whether or not $\alpha>3$ gives a naked singularity is yet to be studied.

\item Throughout the paper, we have considered the possibility of strong singularities which are locally naked. Whether or not they are globally naked is still unknown. The existence of such singularities would be a big blow to the weak cosmic censorship. 
\end{enumerate}

\section{Acknowledgement}

KM would like to acknowledge the support of
the Council of Scientific and Industrial Research (CSIR, India, Ref: 09/919(0031)/2017-EMR-1) for funding the work. KM would also like to thank International Center for Cosmology, Anand, India, for its hospitality.

\end{document}